\documentclass{ws-procs9x6}

\begin{document}

\title{On the detectability of cosmic ray electron spectral features \\
 in the microwave/mm-wave range}

\author{A. TARTARI$^{1*}$, M. GERVASI$^{1,2}$, G. SIRONI$^1$,
M. ZANNONI$^1$ and S. SPINELLI$^1$ }

\address{$^1$ Physics Department "G.Occhialini", University of Milano-Bicocca,\\
$^2$ INFN, Milano-Bicocca,\\
Piazza della Scienza 3, Milano, 20126, Italy\\
$^*$E-mail: andrea.tartari@mib.infn.it\\
www.unimib.it}

\begin{abstract}
Recent measurements of cosmic ray electron energy spectra suggest
that above 10 GeV there may be deviations from a single power law
spectrum. There are hints (ATIC) for a bump occurring between 100
GeV and 1TeV, meaning that there might be more high energy
electrons than expected. Whether these electrons are produced
within pulsar magnetospheres, or due to Dark Matter annihilation
or decay, this is still matter of debate. Understanding the nature
of these ultra high energy particles is a difficult task that can
be fulfilled using all the available astrophysical observables. We
investigate how different energy spectra produce different
observable manifestations in the radio/microwave/mm-wave domain,
where corresponding deviations from a synchrotron power law could
appear. We raise the question around the detectability of these
possible radio spectral features, which may be interesting for a
wide scientific community including astrophysicists and scientists
working on foregrounds removal for CMB experiments.
\end{abstract}


\bodymatter

\section{Science case}

Recently several experiments have pointed out interesting features
in the measured energy spectrum of the $e^+$ and $e^-$ components
of cosmic rays. In particular, we recall here the positron excess
detected by Pamela \cite{pam}\!, the excess of electron counts
($e^+ + e^-$) between 100 GeV and 1 TeV detected by
Fermi\cite{fermi}\!, and the bump in the cosmic ray electron (CRE)
spectrum, centered around $\sim 500$ GeV, detected by
ATIC\cite{atic}\! (not confirmed by H.E.S.S. low energy
analysis\cite{hess})\!. The observational picture will be probably
improved after the observations of AMS-02\cite{ams}\! (which will
be operating in February 2011), due to the large acceptance and
accuracy of the apparatus and to the long operation time (at least
ten years). In this work we do not look for possible
interpretations of these data, rather we study the possibility of
using radio observations to confirm the above CREs spectral
features. In particular, we concentrate on the evaluation of CRE
synchrotron emission in a galactic environment, with
typical values of the magnetic field strength of $1-5$
$\mu$G.

Due to the explorative aims of this study, we limit our attention
to the intensity and degree of polarization of radiation produced
by an ensemble of electrons, in the hypothesis of homogeneity of
the galactic magnetic field on scales greater than the
gyro-radius, and discuss the properties of radiation at the
source, not taking into account radiative transfer through the
interstellar medium (ISM), which affects both intensity and degree
of polarization. This issue can be the object of a subsequent
paper. Here we point out the relevant frequency bands for possible
observations and signatures that may allow one to disentangle
their signal from contaminants like thermal dust emission, peaking
in the sub-mm range. We show also that expected signals are not a
limiting foreground for Cosmic Microwave Background experiments,
being expected at $\sim$ 1 THz and above. Finally we discuss a few
observational aspects of some astrophysical sources which could be
used as targets for this investigation.

\section{Method}

The connection between diffuse radio emission and cosmic ray
propagation within the Galaxy was firmly established in the
Sixties\cite{gs}\!. Here we recall the basic results concerning
the single particle synchrotron emission, and the emission of an
ensemble of particles, with a known energy spectrum, propagating
in a uniform magnetic field \cite{Long94}\!. A charged particle
(an electron, from now on) moving in a region pervaded by a
uniform magnetic field describes an helical trajectory, with a
pitch angle $\alpha$ between the velocity vector and the magnetic
field lines. Being accelerated, it looses energy by radiation. If
the velocity of the charge is relativistic, the radiation is
strongly beamed, and is almost completely emitted within a cone of
aperture $\sim 1/\gamma$, where $\gamma$ is the Lorentz factor,
typically $\geq 10^3$ for the electrons associated to the observed
radio continuum of the Galaxy. This cone is centered around the
instantaneous velocity of the particle, sweeping periodically the
line of sight of the observer. Only in a short ($\sim 1/\gamma$)
fraction of the gyration period, when the line of sight lies
within the radiation cone, the observer registers a pulse of
radiation. The Fourier transform of this pulse, in the observer's
frame, is a continuous spectrum and most of the radiation is
concentrated in a narrow peak at frequency $\nu \sim \gamma^2
\nu_g$, with gyro-frequency $\nu_g=e B/2\pi m_e$, where $B$ is the
magnetic field, $e$ the charge of the electron, and $m_e$ its
mass.

Following Longair, we introduce a cartesian frame suitable for the description of synchrotron
radiation, including its polarization. To this purpose, let's
consider a plane orthogonal to the wave vector {\bf \textsf{k}}: we
can project the magnetic field on this plane. Let's call
$B_{\perp\textsf{k}}$ the projected component. We can form a triad
with a versor parallel to $B_{\perp\textsf{k}}$ (components in this
direction will be labeled with $\parallel$), one orthogonal to both
$B_{\perp\textsf{k}}$ and {\bf \textsf{k}} (components in this
direction will be labeled with $\perp$), and {\bf \textsf{k}}
itself. In this frame, synchrotron emissivities of a single electron
in the two polarizations are: \vskip 0.2cm

$j_{\perp}(\nu) = \Gamma \ B_{\perp} \sin \alpha
\left[F(\nu/\nu_c) + G(\nu/\nu_c) \right]$ \vskip 0.1cm

$j_{\parallel}(\nu) = \Gamma \ B_{\perp} \sin \alpha
\left[F(\nu/\nu_c) - G(\nu/\nu_c) \right]$ \vskip 0.2cm

\noindent where: (1) $\Gamma \cong 1.86 \times 10^{-26}$ W/THz;
(2) $\nu_c = 3 \gamma^2 \nu_g \sin (\alpha/2)$ is the critical
frequency ; (3) $F(x)$ and $G(x)$ are spectral functions defined
through modified Bessel functions as $F(x)=x\int_{x}^{\infty
}K_{5/3}(z)dz$ and $G(x)=xK_{2/3}(x)$. Now, if we have $N(E)dE$
electrons per energy interval $dE$, per unit volume, the spectral
intensity per unit volume radiated by the cloud of electrons is $
I(\nu) = \int_{0}^{\infty}j(x)N(E)dE $, where $x=\nu / \nu_c =
x(E,B)$ and $j=j_{\perp}+j_{\parallel}$. Therefore we arrive at a
useful integral:

\begin{equation}
I(\nu) = \frac{\kappa m_e c^2}{\sqrt{6\nu_g \sin \alpha}}
\sqrt{\nu}\int_{0}^{\infty}j(x)N[E(x)]x^{-3/2}dx \label{aba:eq1}
\end{equation}

\noindent that can be computed analytically in some remarkable
cases (e.g. $N(E) \propto E^{-\beta}$), but that we compute
numerically in order to deal with arbitrary energy distributions.
Here $\kappa$ is normalization factor. An average on the pitch
angle $\alpha$ in the range $(0,\pi)$ will lead to the results
presented in the next section. If we deal separately with the
intensity emitted in two orthogonal polarizations, we can compute
its degree of linear polarization, that is:

\begin{equation}
\Pi(\nu) = \frac{I_{\perp}(\nu)-I_{\parallel}(\nu)}
{I_{\perp}(\nu)+I_{\parallel}(\nu)}. \label{aba:eq2}
\end{equation}

\noindent In terms of Stokes parameters, $\Pi=\sqrt{Q^2+U^2}/I$.
Its value may be up to $\sim 75\%$ for a power law distribution of
electrons with slope around $-3$. Nevertheless this value is
strictly an upper limit, being obtained without considering
radiative transfer effects occurring in the ISM, leading to
depolarization. Conversely, the degree of circular polarization of
an ultra-relativistic electron is vanishingly small, scaling as
$1/\gamma$: therefore, we do not consider $\Pi_c$ in the following
discussion. Like the intensity, the degree of polarization is
computed numerically.

\section{Results}

\begin{figure}[h!]
\begin{center}
\psfig{file=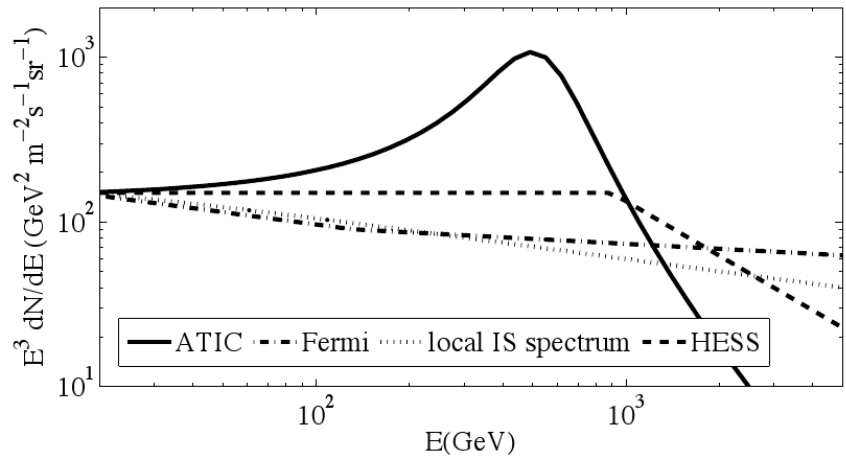, width=4.5in} \caption{Different
analytical description of the electron spectra used for
calculating synchrotron spectra. All these spectra are normalized
at the value of 150 GeV$^2$m$^{-2}$s$^{-1}$sr$^{-1}$ at 10 GeV
energy.} \label{aba:fig1}
\end{center}
\end{figure}

The calculations described in the previous sections (equation
\ref{aba:eq1} and \ref{aba:eq2}), have been implemented using
different electron spectra. In particular we used: (1) the
description of the local interstellar spectrum by Zhang \&
Cheng\cite{lis}\!; (2) a broken power law to describe H.E.S.S.
high energy knee\cite{hess}\!; (3) a synthetic representation of a
Fermi-like spectrum exhibiting a hardening (from a $-3.2$ to a
$-3.0$ slope) around $\sim$ 100 GeV\cite{fermi}\!; (4) a synthetic
representation of an ATIC-like spectrum\cite{atic}\! with a
lorentzian-shaped bump on a constant slope ($\beta=-3$)
background. This oversimplified analytical description is enough
to capture the main features arising in synchrotron emission. All
these spectra have been normalized at a common value of $E^3J(E)$
of 150 GeV$^2$m$^{-2}$s$^{-1}$sr$^{-1}$ at 10 GeV, and are shown
in Fig.\ref{aba:fig1} .

\begin{figure}[h!]
\begin{center}
\psfig{file=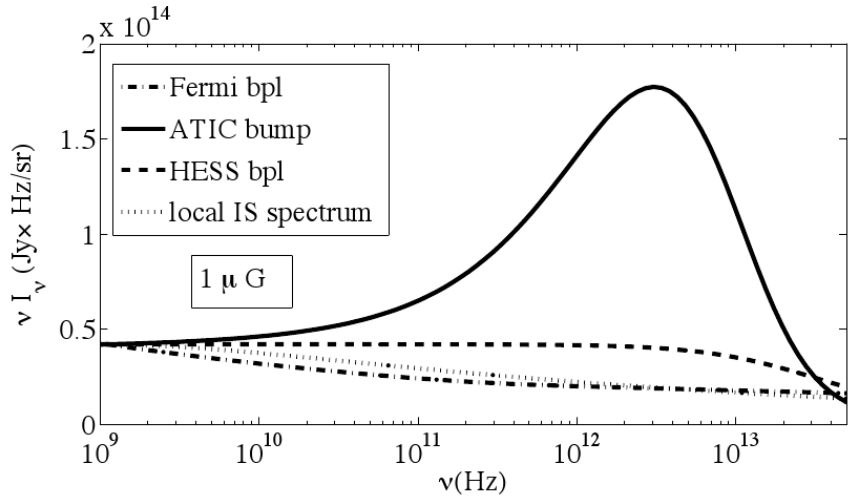, width=4.5in}
\caption{Spectra of the synchrotron emission forecasts for the
different CRE populations. These intensities are normalized to a
radio brightness of $3.8 \times 10^4$ Jy/sr measured by the TRIS
experiment in a purely synchrotron emitting region on the Galactic
halo at $\delta = +42^{\circ}$ and $RA \simeq 10h$. A Galactic
magnetic field of 1$\mu$G is assumed.} \label{aba:fig2}
\end{center}
\end{figure}

\begin{figure}[h!]
\begin{center}
\psfig{file=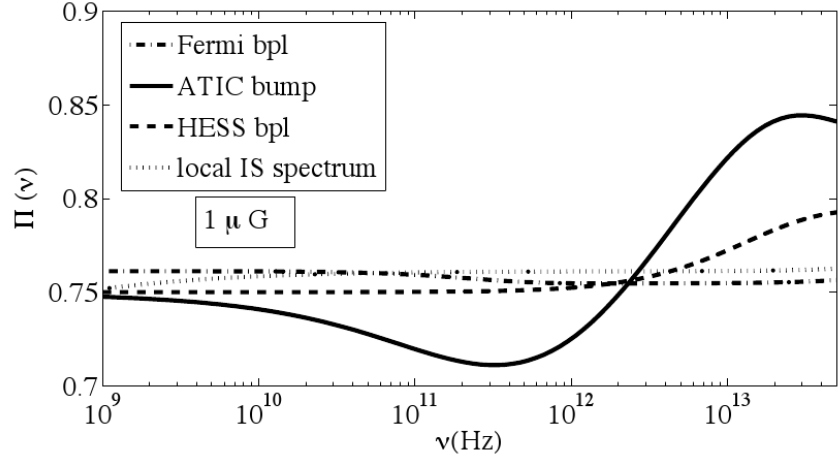, width=4.5in}
\end{center}
\caption{Degree of linear polarization ($\Pi$) {\it{vs}}
frequency. A Galactic magnetic field of 1$\mu$G and the maximum
degree of linear polarization are assumed.} \label{aba:fig3}
\end{figure}

In turn, the synchrotron spectra obtained from these sources are
ultimately normalized to a measured value of the minimum
synchrotron sky brightness at $\delta = +42^{\circ}$ and $RA
\simeq 10h$ (Galactic halo), using the results of the TRIS
experiment presented in Tartari et al. \cite{tris}\!: our
reference brightness at 1 GHz is $3.8 \times 10^4$ Jy/sr. The
obtained synchrotron spectra are shown in Fig.\ref{aba:fig2},
while in Fig.\ref{aba:fig3} we show the degree of linear
polarization.

In Fig.\ref{aba:fig2} and \ref{aba:fig3} we show only the values
obtained for a magnetic field of 1$\mu$G. Results at 3 and 5
$\mu$G are similar except that all the features moves at higher
frequency as expected.

\section{Discussion and Conclusions}

\begin{figure}[h!]
\begin{center}
\psfig{file=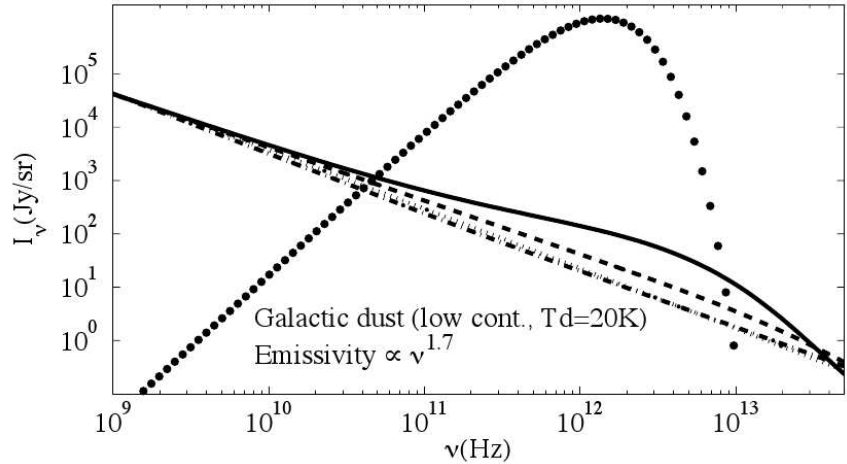, width=4.5in} \caption{Synchrotron
brightness corresponding to different electron spectra (same line
style as in fig.\ref{aba:fig2} and \ref{aba:fig3}.). The filled circles correspond to the emission of cold (20 K) dust in a low contamination region of the sky, normalized to the
measurements of the 100 $\mu$m DIRBE channel.} \label{aba:fig4}
\end{center}
\end{figure}

We have shown that the features in synchrotron intensity and
polarization produced by $>$ 100 GeV electrons fall into the
sub-mm wave/far IR regime (therefore, they are not an issue when
removing foregrounds from CMB maps). Moreover a bump in the
electron spectrum (ATIC) modulates significantly (more than
10$\%$) the degree of synchrotron linear polarization at
frequencies around 1 THz. Unfortunately, the Galactic dust,
through its grey-body emission, $ I_d(\nu) = k_d (\nu /
\nu_0)^{\beta_d} BB(T_d)$, completely dominates the sky brightness
in these bands. In Fig.\ref{aba:fig4} we show the expected
brightness of a clean region of the sky normalized (through $k_d$
emission coefficient) at the DIRBE \cite{dirbe}\! 100 $\mu$m
channel. We assumed a dust temperature $T_d = 20$ K, and grey-body
emissivity scaling as $\nu^{\beta_d}$ ($\beta_d = 1.7$). We see
that dust signal overcomes the synchrotron one by several orders
of magnitude.

\noindent This is just an example, since in general dust
contamination depends on the region observed (different
temperature, density and composition of dust). Let's now consider
different sources of the observed CRE excess.

\emph{A) Dark Matter decay or annihilation in our galaxy.} The
distribution of synchrotron radiation is probably similar to the
expected Dark Matter distribution in the galactic halo. We expect
a smooth distribution with a maximum of intensity in direction of
the galactic center decreasing away from it. The best directions
of observations are far from the galactic disk, where the
synchrotron emission of the background electrons and the thermal
emission of the interstellar dust are faint and more uniform. This
is the situation we have considered to estimate the signals shown
in Fig.\ref{aba:fig4}: above 100 GHz thermal dust emission is the
dominating signal also far from the Galactic disk. Therefore it is
difficult to imagine a detection of a diffuse, smoothly varying
synchrotron signal coming from our own Galaxy, such as what could
be associated to a Dark Matter halo. In this case, also
considering the typical polarization signatures of synchrotron
radiation, the removal of the thermal background would be
extremely difficult, despite the small degree of linear
polarization of thermal dust emission.

\emph{B) Electrons from Pulsar or SNR.} A galactic source like a
pulsar or a SNR can be found (much more easily) on the galactic
disk. Here both the synchrotron emission of the background
electrons and the thermal emission of the interstellar dust are
stronger respect to the signals coming from the halo and shown in
Fig.\ref{aba:fig4}. Dust emission is expected to increase more
than synchrotron. In addition the angular distribution is also
more anisotropic. This situation is compensated by a possible
\emph{boost factor} enhancing the synchrotron signal coming from
these sources, if CRE come out along the direction of observation.
In fact electrons radiate only towards the speed vector, and we
detect their radiation only if this direction is aligned with our
line of sight. Besides electrons originated by these sources come
to the Earth position through a diffusion process, because the
gyro-radius, in the interstellar magnetic field even at energy up
to 1 TeV, is much smaller than the distance of the closest
candidate sources. In addition around these sources the magnetic
field is far from uniform. This \emph{boost factor} could be large
up to the Lorentz factor $\gamma$, but we must take into account
also that the detected CRE could come from sources not radiating
in the direction of our line of sight. In conclusion one could map
a region surrounding a galactic source, like a pulsar or a SNR,
looking for an anisotropic signal at small angular scales. In this
case the background subtraction could be easier, in particular if
we look at the spectral feature in the polarized signal.

\emph{C) Electrons in extragalactic sources.} An alternative
approach is to observe extragalactic radio sources: spiral or
elliptical galaxies. In fact we expect the same phenomenology,
regarding cosmic rays and synchrotron radiation, in external
galaxies as in our own. Also regions surrounding AGNs could be used
as target to observe features in the synchrotron radiation (which
here is much more intense), but our understanding of these sources
is still uncomplete and CRE producing synchrotron radiation are not
observable. \emph{(1)} Spiral galaxies should have synchrotron and
dust emissions similar to the Milky Way, and should have a similar
population of pulsars and SNR in the disk. We can look for an
anisotropic signal against the background at small angular scale,
but the presence of the thermal dust emission does not facilitate
the background removal. In this case we can not decouple the effect
generated by local sources from a Dark Matter signature. \emph{(2)}
Elliptical galaxies show a low dust and gas content. This means that
both thermal emission and SNR generated by supernova explosions in a
low density ISM are no longer important contaminants. Therefore
these galaxies can be used as favorite targets for looking at a
signature of Dark Matter annihilation or decay from the galactic
halo.

In order to investigate the features we pointed out in this paper,
spectral and polarimetric information, together with good angular
resolution, would help. A spectro-polarimeter, or a polarimeter
operating in different photometric bands, installed in the focal
plane of a large far-IR telescope would be necessary. Because of the
frequencies under investigation a space or a balloon-borne
experiment could be preferred.

To have a more realistic estimate of the signals, in all the
situations considered above, detailed calculations have to be
done, including radiative transfer effects, in particular
concerning polarized signals.

\end{document}